\documentstyle[epsf]{article}
\newskip\humongous \humongous=0pt plus 1000pt minus 1000pt

\newif\ifdtup

\catcode`\@=11
 
\@addtoreset{equation}{section}
\def\theequation{\arabic{equation}}
  
\def\@normalsize{\@setsize\normalsize{15pt}\xiipt\@xiipt
\abovedisplayskip 14pt plus3pt minus3pt%
\belowdisplayskip \abovedisplayskip
\abovedisplayshortskip \z@ plus3pt%
\belowdisplayshortskip 7pt plus3.5pt minus0pt}
 
\def\small{\@setsize\small{13.6pt}\xipt\@xipt
\abovedisplayskip 13pt plus3pt minus3pt%
\belowdisplayskip \abovedisplayskip
\abovedisplayshortskip \z@ plus3pt%
\belowdisplayshortskip 7pt plus3.5pt minus0pt
\def\@listi{\parsep 4.5pt plus 2pt minus 1pt
     \itemsep \parsep
     \topsep 9pt plus 3pt minus 3pt}}
 
\relax

\catcode`@=12
 
\catcode`\@=11
 
\def\section{\@startsection{section}{1}{\z@}{3.5ex plus 1ex minus
   .2ex}{2.3ex plus .2ex}{\large\bf}}
 
\def\thesection{\arabic{section}.}

\def\appendix{\setcounter{section}{0}
 \def\thesection{Appendix \Alph{section}:}
 \def\theequation{\Alph{section}.\arabic{equation}}}
\begin{document}
\begin{titlepage}
\begin{center}
{\LARGE
 Confinement, Supersymmetry Breaking   and  $\theta$ Parameter Dependence in the
 Seiberg-Witten Model} \end{center}

\vspace{1em}
\begin{center}
{\large
K. Konishi}
\end{center}

\vspace{1em}
\begin{center}
{\it {\large 
 Dipartimento di Fisica -- Universit\`a di Genova\\
     Istituto Nazionale di Fisica Nucleare -- sez. di Genova\\
     Via Dodecaneso, 33 -- 16146 Genova (Italy)\\
     E-mail:  konishi@ge.infn.it \\}}
\end{center}

\vspace{7em}
{\bf ABSTRACT:}
{\large   By using the standard perturbation theory we study the mass as well as
$\theta$ parameter dependence  of the  
Seiberg-Witten theory
 with $SU(2)$ gauge group, supplemented with a $N=1$ supersymmetric
 as well as  a
smaller nonsupersymmetric gaugino mass.    Confinement persists at
all values of the bare  vacuum parameter $\theta_{ph}$; 
as it is varied there  is 
however a
phase transition at  $\theta_{ph} =-\pi, \pi, 3\pi, \ldots$.  At 
these values of $\theta_{ph}$ the vacuum is doubly degenerate and CP
invariance is spontaneously broken \`a la Dashen. 
Due to the instanton-induced renormalization
effect the low energy effective $\theta$ parameter 
remains small irrespectively of 
$\theta_{ph}$  as long as the gaugino mass is sufficiently small.    }

\vspace{15em}
\begin{flushleft} 
{\large Talk presented at
Int. Conf. on  High Energy Physics  (Warsaw, Poland)   July
1996} \end{flushleft} 
\end{titlepage}

\def\beq{\begin{equation}}
\def\eeq{\end{equation}}
\def\non{\nonumber}
\def\De{\Delta}
\def\bea{\begin{eqnarray}}
\def\eea{\end{eqnarray}}
\def\bra{\langle}
\def\ket{\rangle}
\def\de{\partial}
\def\si{\sigma}
\def\sb{{\bar \sigma}}
\def\rn{{\bf R}^n}
\def\r4{{\bf R}^4}
\def\s4{{\bf S}^4}
\def\Tr{\hbox{\rm Tr}}
\def\ker{\hbox{\rm ker}}
\def\dim{\hbox{\rm dim}}
\def\sup{\hbox{\rm sup}}
\def\inf{\hbox{\rm inf}}
\def\re{\hbox{\rm Re}}
\def\im{\hbox{\rm Im}}
\def\infi{\infty}
\def\nrm{\parallel}
\def\nrmi{\parallel_\infty}
\def\o{\over}
\def\teo{\noindent{\bf Theorem}\ }
\def\all{\hbox{\rm all}}
\def\co{$^,$}
\def\daa{$^-$}
\def\dirac{{\cal D}}
\def\dplus{{\cal D_{+}}}
\def\dminus{{\cal D_{-}}}
\def\om{\Omega}
\def\cost{\hbox {\rm cost.}}
\def\o{\over}
\def\th{\theta}
%%%%%%%%Substitutions mld=m_{\lambda};lmd=\lambda; Lmd=\Lambda

In  celebrated papers, Seiberg and Witten have solved exactly
an $N=2$ supersymmetric non-Abelian geuge theory in four dimensions,
determining  the vacuum degeneracy, and in each vacuum, the exact spectrum
and  interactions among  light  particles.  An especially 
interesting observation of Ref \cite{SW}, made in the  pure $N=2$
supersymmetric Yang-Mills theory with $SU(2)$ gauge group, 
 is that upon turning 
on the mass term\footnote{We follow the notation of Ref\cite{SW}.}
 $\int d^2\theta \,Tr\, m\Phi^2, $   $m
\ll \Lambda, $   the light  magnetic monopole field
condenses,  providing thus the first explicit  realization of the confinement
mechanism envisaged by  't Hooft.\cite{TH} 

In our  opinion, the  solution    of Seiberg and Witten  plays a role 
 in the context of four dimensional gauge
theories with strong interactions, similar to the one  played by the 2 dimensional
Ising model in  statistical physics,   in   that it exhibits  in a context of an
exactly solved model the kind of phenomena which can occur in  more realistic models
(such as QCD, in our case).

We  shall focus our attention in this talk on several questions related to
confinement, mass dependence, supersymmetry breaking  and the dependence of
physics on the CP violation parameter. 

   The  model is defined by the Lagrangian,
 \beq  {\cal L} =  {1\over 4\pi} \im \, \tau_{cl} \left[\int d^4 \th \,
\Phi^{\dagger} e^V \Phi +\int d^2 \th\,{1\o 2} W W\right] + \int d^2 \th \,m
\,\Phi^2 + h.c. +   \Delta {\cal L}, \eeq
where 
\beq \tau_{cl}={\th_0 \o 2\pi} + {4 \pi i \o g_0^2}, \label{struc}\eeq
and the supersymmmetry breaking term $\Delta {\cal L}$ can be taken as e.g., 
\beq  m_{\lambda}\lambda \lambda +h.c, \,\, m^{'2}\phi^2 + h.c., \,\, 
|m^{''}\phi|^2,  \,\, {\hbox{\rm or}}\,\,  m^{''} \psi^2 + h.c. 
\eeq
with susy breaking mass much smaller than $m$, e.g., 
\beq   m_{\lambda} \ll  m \ll \Lambda. \eeq
$\Phi= \phi + \sqrt2 \th \psi +\ldots,$ and $W_{\alpha} = -i \lambda + {i \o 2}
(\si^{\mu} \sb^{\nu})_{\alpha}^{\beta}  F_{\mu \nu} \th_{\beta} +\ldots $ are both
in the adjoint representation of the gauge group.  In the massless case 
($m_{\lambda}=m=0.$) the theory has a global symmetry, $SU_R(2) \times Z_8$. 
Under the $SU_R(2)$ the fermions $(\lambda, \psi)$ transform as a doublet, while
the bosons $(A_{\mu}, \phi)$ are singlets. $Z_8$ is an anomaly-free remnant of 
$U_{\Phi}(1)$. 

The case of susy breaking with $ m^{'2}\phi^2 $ is entirely similar to the gaugino
mass case discussed below; the case with $|m^{''}\phi|^2$ or $ m^{''} \psi^2$
is not very interesting since no $\theta$ dependence is introduced. For these
reasons  we shall discuss the case with gaugino mass in this talk.

\bigskip
 \leftline {\bf 1.  Massless (N=2)  theory   }
This is the case exactly solved by Seiberg and Witten \cite{SW}.
Classically the theory has a flat direction along $\phi = a \tau_3$, with $a$ an
arbitrary complex number. Quantum mechanic1ally if $u \equiv \bra \Phi^2 \ket \ne 0$
Higgs phenomenon occurs, the $SU(2)$ gauge group is broken to a $U(1)$ subgroup, 
and the low energy degrees of freedom are $U(1)$ gauge supermultiplet $W$ and 
its $N=2$ susy partner scalar multiplet $A$.  They are described by an effective
Lagrangian (whose form is dictated by the $N=2$ supersymmmetry)
\beq  L_{eff} = {1\over 4\pi} Im \, [ \int d^4\theta \, {\de F(A) \over \de A} {\bar
A} +  \int d^2 \theta {1\over 2} {\de^2 F(A) \over \de A^2} WW ], \eeq
where $F(A)$, holomorphic in $A$,  is called prepotential. As seen from this
expression (the vev of ) $ {\de^2 F(A) \over \de A^2} $ plays the role of the
effective $\theta$ parameter  and coupling constant, 
$\tau(a)= {\de^2 F(a) \over \de a^2}=\theta_{eff}/2\pi+4\pi i/g_{eff}^2.$
Perturbative and nonperturbative (instanton) corrections lead to the general form
$$F(a)={i \over 2\pi}A^2 \log {A^2\over \Lambda^2} + F^{inst},$$
where $ F^{inst}=\sum_k(\Lambda/A)^{4k} A^2$ is the contribution from
multiinstanton effects. 

By exploiting  some earlier results Seiberg and  Witten manage to find the exact
mass formula for the dyons of the theory
  \beq  M_{n_e, n_m} =\sqrt2 |Z|, \quad Z=a
n_e + a_D n_m \label{massfor}\eeq
 where $n_m$ and $n_e$ are integer-valued  magnetic and electric quantum numbers
labelling the particles.

The remaining problem of determining the dependence of $a(u)$ and $a_D(u)$ on 
$u \equiv \bra \Phi^2 \ket $ has been solved by studying the monodromy properties 
(how $a(u)$ and $a_D(u)$  transform under various  closed path transformations $u \to
u$) and an Ansatz of massless monopole at $u=\Lambda^2$.       The solution is
\beq a={\sqrt2\over \pi}\int_{-\Lambda^2}^{\Lambda^2} dx\, {\sqrt{x-u}\over
\sqrt{x^2-\Lambda^4}} = {\sqrt2 \Lambda \over \pi}\int_{-1}^{1} dx\,
{\sqrt{x-u/\Lambda^2}\over \sqrt{x^2-1}};\eeq
\beq 
 a_D={\sqrt2\over \pi}\int_{\Lambda^2}^{u} dx\, {\sqrt{x-u}\over
\sqrt{x^2-\Lambda^4}} = {\sqrt2 \Lambda \over \pi}\int_{1}^{u/\Lambda^2} dx\,
{\sqrt{x-u/\Lambda^2}\over \sqrt{x^2-1}} \label{swsol}\eeq
(they can be expressed by hypergeometric functions). 
This  solution  amounts to the calculation of the full spectrum
(Eq.(\ref{massfor})) and  an exactly
known low energy effective action  including  all multi-instanton
corrections.

It is  well known ("Witten  effect") that the electric charge of the dyon in a 
 non abelian theory (spontaneously
broken to a $U(1)$ theory) is given, in the  $\theta $ vacuum, by
$$q = n_e + {\theta\over 2\pi}n_m,$$
where $n_e$ $n_m$ are integer quantum numbers.  Due to this effect  the exact
periodicity in $\theta_{eff}$ is guaranteed in spite of a nontrivial  spectral
flow  when $u$ goes from $\Lambda^2 $ to $- \Lambda^2 $ and back, ecc. 

Although one expects no dependence of physics  on the bare $\theta$ parameter 
in the massless theory,  the way such an independence is realized is quite
nontrivial. A shift of $\theta$ by $\Delta \theta$ causes the change in $ \Lambda$ 
as  
$ \Lambda \to  e^{i \Delta \theta/4} \Lambda. $
But if we now move to a different vacuum by
$  u  \to e^{i \Delta \theta/2} u $
 the net change is 
\beq    a \to e^{i \Delta \theta/4} a;\quad a_D \to e^{i \Delta \theta/4} a_D:
\label{rotaead}\eeq
a common phase rotation of $a$ and $a_D$. 
Thus all physical properties of an appropriately shifted vacuum $u$ with a new 
value of $\theta$  are the same as in the original theory.
     In other  words,
the ensemble of theories  represented by the points of QMS,
 taken together,  is
invariant under $\theta \to \theta + \Delta \theta.$ 
Vice versa, at fixed generic  $u$
physics  depends on $\theta$  non trivially, since 
  CP invariance is spontaneously broken by $u\ne 0$.

(The above argument may be inverted to the statement  that  the   anomalous chiral
$U(1)$ transformation property of the theory is indeed correctly incorporated in the
low energy action.)

The effective Lagrangian near $u= \Lambda^2$  contains  the magnetic monopole fields
$(M, {\tilde M})$:
 \bea &&{\cal L}^{(\Lambda^2)}= {1\over 4\pi} Im
\, [ \int d^4\theta \, {(\de F(A_D)/\de A_D)} {\bar A_D} +  \int d^2 \theta \,
{(\de^2 F(A_D)/ \de A_D^2)} W_D W_D/2 \,] \non \\
  &+&  \int d^4\theta \,[ M^{\dagger}e^{V_D} M + 
{\tilde M}^{\dagger}e^{-V_D} {\tilde M}] + 
\int d^2 \theta \,   \sqrt{2} A_D {\tilde M} M + h.c. \label{lagmono}\eea
where the particular Yukawa interaction $\sqrt{2} A_D {\tilde M} M$ is consistent
with the mass formula Eq.(\ref{massfor}) and is related by  $N=2$ supersymmetry
to the monopole kinetic terms. In Eq.(\ref{lagmono}) the fields
$(\lambda_D, \psi_D)$ and $(M, {\tilde M}^{\dagger})$ form doublets of $SU_R(2)$,
while other  components are singlets. 

\bigskip
 \leftline{\bf  $N=1$ theory: $m\ne 0; m_{\lambda}=0$.}
Consider now  adding  a mass term, $m \Phi^2|_F \equiv mU|_F $ in the fundamental
action.  It  reduces  the $N=2$ supersymmetry to $N=1$ supersymmetry.  
Near the point $u=\Lambda^2$ where  the $(1,0)$ magnetic monopole becomes massless
the effective low energy action 
 is given by Eq.(\ref{lagmono}), implemented by the superpotential 
$m \Phi^2|_F \equiv mU|_F. $
Extremizing
the full superpotential 
$   \sqrt2 A_D {\tilde M}M + mU(A_D)   $
with respect to $A_D$ and $M$, yields
\beq A_D=0, \quad i.e., \,\, u= \Lambda^2   \eeq  
(so that keeping only $(1,0)$
particles besides $(A_D, W_D)$ is consistent)  and the magnetic monopole
condensation, \beq 
  \bra M \ket =  \bra
{\tilde M} \ket= (-{m \,U^{'}(0)/\sqrt2})^{1/2}.  \label{magncond}\eeq 
The theory thus  confines the color electric charge, \`a la 't Hooft; at the same
time the value of condensate $u= \bra \Tr \, \Phi^2 \ket$ is now fixed at
$\Lambda^2$.

Actually,  an analogous thing happens if one starts with a theory near $u=
 - \Lambda^2$  (at which the $(1,1)$ "dyons"  $N, {\tilde N}$ become massless). The 
low energy Lagrangian has the same structure as (\ref{lagmono}), except for
the replacement,
\beq  M, {\tilde M} \to N, {\tilde N}; \quad  A_D \to A_D^{'}\equiv A_D+A. \eeq
Upon the addition of the mass term, the $N$ field condenses, as in
Eq.(\ref{magncond}), where $U^{'}(0)$ now refers to the derivative and the value
($A_D^{'}=0$)    with respect to $A_D^{'}$. Also, by the Witten effect, the $N$
particle becomes a pure magnetic monopole, with zero electric charge at 
$u= - \Lambda^2$ (where $\theta_{eff}= - 2 \pi$.)

One thus finds that the continuous degeneracy of the vacua in the $N=2$ theory is 
 (almost) eliminated    by the mass perturbation,  leaving only  a double degeneracy
$ u = \pm \Lambda^2.$  This is of course the right number of the vacua
 (Witten's index) in a 
massive $N=1 $ supersymmetric $SU(2)$ gauge theory. 
  
As is usual in the  standard degenerate perturbation theory in quantum mechanics,
to lowest order  the only effect of the perturbation is to  fix the vacuum to the
"right",    but  unperturbed,   one, with $N=2$ (hence $SU_R(2)$) symmetric
properties.  This means that  the  $SU_R(2)$ current of the original
theory, \beq   J_{\mu}^a = \Tr \,{\bar \Psi} {\bar \sigma_{\mu}} \tau^a \Psi
\label{currenthigh}\eeq
 where $\Psi = \pmatrix{\lambda \cr \psi},$   is
well approximated  in the low energy effective theory by the Noether currents
(of the low energy theory) 
  \beq J_{\mu}^a = {\bar \Psi_D} \,{\bar
\sigma_{\mu}} \tau^a \Psi_D + i (D_{\mu} S^{\dagger})  \tau^a  S -
i  S^{\dagger}  \tau^a  D_{\mu} S,
\label{currentlow}\eeq  
where $\Psi_D = \pmatrix{\lambda_D \cr \psi_D}$  is the dual of the (color diagonal 
part of)  $\Psi$,  and $S\equiv \pmatrix { M \cr {\tilde M}^{\dagger}}$.

For instance an  (appropriately normalized) $R$-like ratio, 
$  R= Disc_{q^2} \Pi(q^2),  $
associated with  the current-current correlation function 
\beq i \int d^4 x \, e^{-iqx} \bra 0| T\{ J_{\mu}^3(x) J_{\nu}^3(0)\}|0\ket
= (q_{\mu}q_{\nu} - q^2 g_{\mu\nu} ) \Pi(q^2), \label{correl}\eeq
is given at high energies by (asymptotic freedom): 
\beq R_{q^2 \gg \Lambda^2} \simeq  6 +  O(\alpha(q^2),
q^4/\Lambda^4).    \label{af}\eeq
At low energies, $R$ counts  the weakly interacting
dual particles  and  magnetic monopoles,
\beq R_{m\Lambda  \ll q^2 \ll \Lambda^2} \simeq  3. \eeq
 This amounts to
an exact resummation of an infinite number of power and  logarithmic corrections
in Eq.(\ref{af}). 

To next order the effect of the explicit  $SU_R(2)$  breaking,
\beq \de^{\mu} J_{\mu}^{-} = i\, m \, \Tr \, { \lambda }{\psi};\quad 
\de^{\mu} J_{\mu}^{3}=  {i\over 2}( \Tr \, m \psi^2 - h.c. )\eeq
 must be
taken into account.\footnote{Although we are discussing here the first order mass
corrections  in a degenerate perturbation theory,  many  crucial relations
are  exact due to the  nonrenormalization theorem (the form of the
superpotential, etc). Most results  below, then  survive higher order 
corrections which affect only   D   terms.}  
In particular,     the  anomaly of Ref \cite{KK} shows that 
\beq   m\,u = m\, \bra  \Tr \,\phi^2 \ket  = 2 {g^2\over 32\pi}\bra \Tr \,
\lambda(x) \lambda(x)\ket, \label{konishi}\eeq  hence 
\beq {(g^2/ 32\pi)}\bra \Tr \,\lambda(x) \lambda(x)\ket = \pm m \Lambda^2/2. \eeq 
 Also,  the 
supersymmetric Ward-Takahashi  like identity 
$  \bra \Tr \, \psi^2 \ket =2 \,m^{*}
\bra  \Tr \,\phi^{*} \phi  \ket, $
 and the fact that $\phi$ is in the
adjoint representation hence probably $ \bra  \Tr \,\phi^{*} \phi  \ket \sim \cost
\bra  \Tr \,\phi^2  \ket \simeq \Lambda^2,$ suggests that $  \bra \Tr \,\psi^2 \ket
\sim O(m\Lambda^2 )$ also. 

Due to the magnetic monopole condensation the $SU_R(2)$ current produces 
the light particles with strength
\beq   F   \sim  \bra M \ket =O(m^{1/2}
\Lambda^{1/2}). \label{fpi}\eeq 
Also, it is easy to see that those are  (to lowest order) the real and imaginary parts of $M -
{\tilde M}$ and the imaginary part of  $  \bra M^{\dagger}\ket (M + {\tilde
M})$, apart from normalization.   
(Actually,  a linear combination of the real and imaginary parts of 
$M - {\tilde M}$ becomes the longitudinal part of the dual vector boson by the Higgs
mechanism.) 

The masses of the   particles  can be  obtained 
most easily  from the fermion bilinear terms arising from the Yukawa interaction
terms upon shifting the magnetic monopole fields  (Eq.(\ref{magncond})).  They are
all of order of  $ \sim O(m^{1/2}\Lambda^{1/2}/ \log^{1/2}(\Lambda/m))$.

In spite of the fact that 
the light
particles  have mass,
\newline
 $ \sim O(m^{1/2}\Lambda^{1/2}/ \log^{1/2}(\Lambda/m)) \ll \Lambda ,
$ and  $F \ne 0$,   these bosons cannot be interpreted quite as 
pseudo Goldstone bosons.\footnote{We thank M. Testa for discussions on this  issue.}
  If one attempts to write the Dashen like  formula\cite{RD} 
by saturating the current-current
correlation functions  such as Eq.(\ref{correl}) (at $q^2 \le m \Lambda$)
 by the lowest
lying particles, one finds that the pole term does not actually dominates. 
The usual pole enhancement as compared to the continuum contribution (by a power
of  $\Lambda/m $ ) is here compensated by the suppression  $F_{\pi}^2 =O(m)$, hence 
 the pole term and continuum contribute both as $O(m^2)$, and are  
of the same order as the right hand side
$ m \bra \Tr \,\psi^2 \ket + h.c.$  which is also one power of $m$ down 
as compared the standard Dashen's formula, due to 
the supersymmetric Ward-Takahashi like 
relation.
Also, from the Yukawa Lagrangian one sees that the particles ($A_D, W_D$)  become
massive only through mixing with the magnetic monopoles. Since $N=1$ theory
 is expected 
to confine, with all particles massive,    the two  $N=1$ vacua could not be
anywhere else in the QMS of $N=2$ theory,  except at the two singular points $u=\pm
\Lambda^2$.     
Finally, CP invariance is  exact in the $N=1$ theory:\cite{DPK}  the theory is independent of
the value of the bare $\theta$ parameter as well as of  $\arg m,$, furthermore spontaneous 
CP breaking does not take place  in spite of a non
zero vacuum expectation values $u=\bra \Tr \Phi^2 \ket$ and   $\bra M \ket$.  This is the
consistent with  the uniqueness
  (except for discrete, double degeneracy) of the vacuum.
  
\bigskip

\leftline {\bf Supersymmetry breaking}

\begin{figure}[h]
\begin{center}
\leavevmode
\epsfxsize=8cm
\epsffile{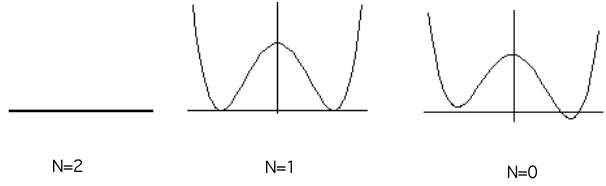}
\end{center}
\caption{Ground states in $N=2$, $N=1$ and in the $N=0$ theories.}\label{Vacua}
\end{figure}

\begin{figure}[h]
\begin{center}
\leavevmode
\epsfxsize=10cm
\epsffile{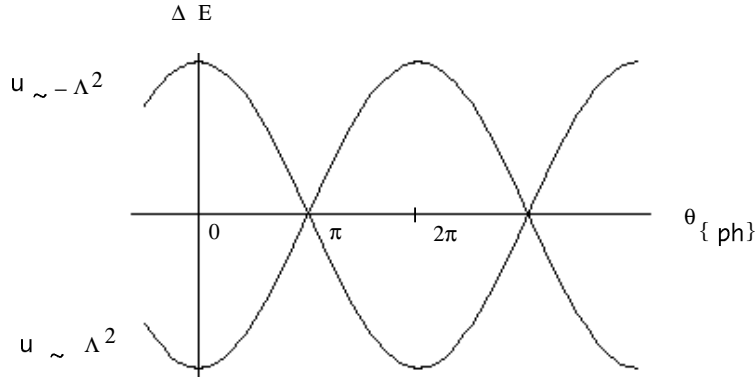}
\end{center}
\caption{Energy density in the two minima}\label{Cosine}
\end{figure}

Consider now adding    the gaugino mass such that $0 < m_{\lambda} \ll m \ll \Lambda.
$  It is important that  the  $N=1$ theory with $m \ne 0$ has two well-defined
vacua: working in one of them the addition of
the gaugino mass term  can be treated by the standard perturbation theory. The 
shift of energy is then given by
 \beq   \Delta E = - \bra m_{\lambda} \lambda \lambda  \ket  + h.c.
= - ( 16 \pi^2 m_{\lambda}  m / g^2) \bra \Tr \Phi^2 \ket  +h.c. = \mp 2 \epsilon
\cos \theta_{ph}/2,  \label{shiften}\eeq
where 
\beq \theta_{ph}\equiv \theta + 2 (\arg m + \arg m_{\lambda}), \eeq
$\epsilon= (16 \pi  / g^2) |m_{\lambda} \, m \, \Lambda^2|$ and the anomaly
Eq.(\ref{konishi})  has been used.
 Note that $\theta_{ph}$ is the correct  (bare) vacuum  parameter 
 on which physics  depends. 
 Eq.(\ref{shiften}), corresponding to the
lower  energy value,  represents the physical energy density of the vacuum,
since the rest of the potential gives a vanishing contribution due to 
supersymmetry.\footnote{A  result somewhat analogous 
to (\ref{shiften}) had earlier been obtained 
independently   by Evans, Hsu and others also\cite{Hsu}. 
We thank Evans and Hsu for discussions on their results prior to publication.
Soft supersymmetry breaking in the context of Seiberg-Witten theory 
was studied also in \cite{SBR}.} 
 Eq.(\ref{shiften}) shows that  the remaining double degeneracy of $N=1$ vacua
has been  lifted: the  ground state  is unique and is  near  $u= +\Lambda^2$ for $0 <
\theta_{ph} <\pi$ and near  $u= -\Lambda^2$  for $\pi < \theta_{ph} < 2\pi$. 
See Fig.\ref{Vacua}, Fig.\ref{Cosine}. 

 Note that 
$\Delta E$ is periodic in  $\theta_{ph}$ with period $2\pi$ in spite of the
appearance of a cosine with  half angle. 
At precisely $\theta_{ph}=\pi$ 
 the two vacua  are exactly
degenerate:  CP  symmetry 
(which exchanges the two vacua) is spontaneously broken
(Dashen's mechanism).  As $\theta_{ph}$  crosses the value  $\pi$  a phase
transition occurs. At the both sides there are 
similar-looking confinement phases with
dyon condensation,  but the fields which condense at  
 $\theta_{ph} < \pi $ and at $\theta_{ph} > \pi $ are different
 and relatively nonlocal so that there is a nontivial
rearrangement of the vacuum. 
It is instructive  to study  the  "wave function corrections", by minimizing the
scalar potential  resulting from the Lagrangian,
\beq {\cal L}^{(\Lambda^2)} + m \, \Tr \, \Phi^2 |_F + h.c. + \epsilon U(A_D)|_0 +
h.c.\eeq 
where $|_0$ indicates the lowest component of the superfield. By using the expansion
$   U(A_D)=\Lambda^2 (1  + f({A_D/i\Lambda}) ); 
\quad   f(x)= 2x + {(1/4)}x^2 -{(1/
32)} x^3 +\ldots,$   obtained by the inversion of the Seiberg-Witten solution
(\ref{swsol}), one
finds that the vacuum is slightly shifted from $a_D=0, \, u = +\Lambda^2.$  To
lowest order in $m_{\lambda}$ one finds 
$a_D/i \Lambda= (4 \sqrt2 \pi^2/g^2) \cdot |m_{\lambda} / \Lambda| e^{-i
\theta_{ph}/2};$ therefore 
\beq u\simeq \Lambda^2 \{ 1 +\eta \,
e^{-i \theta_{ph}/2}\},   \label{vacuum1}\eeq
where
 \beq   \eta \equiv 
(8 \sqrt2 \pi^2/g^2) \cdot |m_{\lambda} /
\Lambda|  \ll 1.  \eeq  
Thus for $-\pi <\theta_{ph} < \pi$  (where it is the true vacuum)  the vacuum near
$u=\Lambda^2$  is shifted slightly towards right in the $u/\Lambda^2$ plane, while
for  $\pi <\theta_{ph} < 3\pi$  (where it is only a local minimum) it is shifted
towards left. At $\theta_{ph}=\pi$,   $u$ crosses the line ${\rm Arg}
u= -i\pi/2$.

Similarly,  the other vacuum (near $u=e^{i\pi} \Lambda^2$) is found to be at  
\beq u\simeq \Lambda^2 \{ - 1 + \eta \,
e^{-i \theta_{ph}/2}\}.\eeq 

These results are interesting because    the spectrum of the
 $N=2$ theory is known to change  discontinuously along a closed   curve in the
space of $u$ (which is an almost  ellipse and passes  the points $u= \pm \Lambda^2$
vertically, see Ref\cite{SW, FB}).  Thus the vacuum of the massive theory is
always in a  region (called weak-coupling region in \cite{FB}) which is smoothly
connected to the semiclassical ($u \to \infty$) limit.

\begin{figure}[h]
\begin{center}
\leavevmode
\epsfxsize=8cm
\epsffile{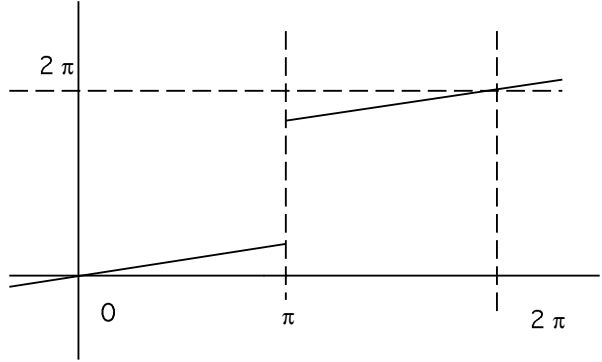}
\end{center}
\caption{$\theta_{eff}$ vs $\theta_{ph}$}\label{theta}
\end{figure}

The dependence of the low energy effective  theta
parameter $\theta_{eff}= 2\pi \re da_D/da $ on   $\theta_{ph}$
can be easily found from the above. 
One finds that for $-\pi < \theta_{ph} < \pi$
\beq    {\theta_{eff} \o 2\pi} \simeq {\pi \theta_{ph} \o  2 \log^2 (1/\eta)} -
{\pi \eta \sin (\theta_{ph}/2) \o 8 \log (1/\eta)},\eeq
 while   for $-3\pi < \theta_{ph} < - \pi$
\beq    {\theta_{eff} \o 2\pi} \simeq   
  -1 + {\pi (\theta_{ph}+ 2\pi)  \o  2 \log^2 (1/\eta)} +
{\pi \eta \sin (\theta_{ph}/2) \o 8 \log (1/\eta)}, \eeq
when $\eta \ll 1$.    These results are illustrated in Fig.\ref{theta}.  Again, $\theta_{eff}$
   (which is related to the physical electric charge of the monopole) 
   depends on the physical combination $\theta_{ph},$ as it should. 
   
One thus sees that irrespective of the value of the bare theta parameter
the low energy  electric and magnetic couplings $g_{eff} \theta_{eff} < g_D \sim 1/\log
^{1/2}(\Lambda/m_{\lambda})$
of the magnetic monopole both remain small as long as $m_{\lambda}/\Lambda$ is sufficiently
small. The weak coupling description of the low energy physics is
 valid at all $\theta_{ph}$:
oblique confinement \`a la 't Hooft\cite{TH}  does not occur at
$\theta_{ph} = \pi$ in this model.  With a larger supersymmetry 
breaking (e.g.,QCD), however, 
the question of oblique confinement remains open.

On the other hand, confinement persists at all bare $\theta$ parameter,
providing thus an explicit counterexample to the 
conjecture of Schierholtz\cite{Schier}.

In keeping the low energy effective theta parameter small the effect of renormalization of the $\theta$ parameter in the
infrared due to multiinstantons (dyon loops) is crucial, an aspect  little understood
in a more realistic theory of interest (QCD). This is a kind of   phenomenon alluded
in the beginning of the talk, which  seems to deserve a deeper study.

\end{document}